# An improved model for describing the net carrier recombination rate in semiconductor devices

M. L. Inche Ibrahim[1,*] and Anvar A. Zakhidov[2,3]

[1]Department of Science in Engineering, Faculty of Engineering, International Islamic University Malaysia, Kuala Lumpur, Malaysia

[2]Physics Department and NanoTech Institute, University of Texas at Dallas, Richardson, TX 75083, USA

[3]ITMO University, St. Petersburg, Russia

* e-mails: lukmanibrahim@iium.edu.my, mlukmanibrahim@gmail.com

**Abstract**

Carrier recombination is a process that significantly influences the performance of semiconductor devices such as solar cells, photodiodes, and light-emitting diodes (LEDs). Therefore, a model that can accurately describe and quantify the net carrier recombination rate in semiconductor devices is important in order to further improve the performance of relevant semiconductor devices. The conventional model for describing the net carrier recombination rate is derived based on the condition that there is no electric current in the considered semiconductor, which is true only when the semiconductor is not part of a semiconductor device, and hence is not connected to an external circuit. The conventional model is adopted and used for describing the net carrier recombination rate in semiconductors that are part of devices (i.e. in semiconductor devices). In this paper, we derive and propose a new model for describing the net carrier recombination rate in semiconductor devices. The newly proposed model is an improvement to the currently used model by considering the fact that electric current can flow in the semiconducting materials of semiconductor devices. We validate the proposed recombination model and show that the use of the proposed model can be crucial for modeling and analyzing the performance of optoelectronic devices such as solar cells and LEDs.

**1        Introduction**

Recombination of electrons and holes is a fundamental process that occurs in a semiconductor whether the semiconductor is part of a semiconductor device or not. Carrier recombination is of special interest to several semiconductor devices. In solar cells and photodiodes, all carrier recombination processes should be minimized, whereas in light-emitting diodes (LEDs), radiative recombination should be maximized, and other possible recombination processes should be minimized. Therefore, an accurate description and quantification of the carrier recombination rate is important to accurately identify and evaluate the possible areas of improvement for those devices. An accurate quantification of the carrier recombination rate could also be important for studies on the fundamental properties of semiconductors.

For a given carrier recombination process, the net carrier recombination rate per unit volume $R_{\text{net}}$ in a semiconductor is given by [1, pp. 313-324]

$$R_{\text{net}} = R_{\text{total}} - G_0, \qquad (1)$$

where $R_{\text{total}}$ is the total carrier recombination rate per unit volume in the semiconductor, and $G_0$ is the carrier generation rate per unit volume that inherently or permanently occurs in the semiconductor. In a semiconductor that is not part of a device (called an isolated semiconductor here), it is well known that $G_0$ originates from thermal agitation, which causes electrons in the valence band jumping either directly or indirectly into the conduction band, thus producing electrons in the conduction band and electron vacancies (holes) in the valence band [1, pp. 313-324]. In a semiconductor that is part of a device, there is an additional factor that contributes to $G_0$ as will be shown in this paper.

The conventional model for describing the detail expression of $R_{net}$ is derived under the condition that there is no electric current in the considered semiconductor, which is true only when the semiconductor is not part of a device and not connected to an external circuit. However, this conventional model is adopted and widely used to describe $R_{net}$ even in semiconductors that are part of devices (i.e., in semiconductor devices). In this paper, we derive a new model for describing $R_{net}$ in semiconductor devices and show that it should be different compared with the conventional $R_{net}$ model. The newly proposed model is an improvement to the currently used model by considering the fact that electric current can flow in semiconductors that are part of devices. We validate the proposed model and show that the use of the proposed model is important to better analyze and enhance the performance of relevant semiconductor devices.

This paper is organized as follows. In Sec. 2, we describe the carrier continuity equations and the carrier recombination in semiconductors. In Sec. 3, first we rederive the conventional model for quantifying $R_{net}$, and then we derive a new improved model for quantifying $R_{net}$ in semiconductor devices. In Sec. 4, we employ organic solar cells (OSCs) as a case study to validate and demonstrate the significance of the newly proposed model. In Sec. 4, we also provide some physical insights regarding the newly proposed $R_{net}$ model. The paper is then concluded in Sec. 5.

## 2 Carrier continuity equations and carrier recombination in semiconductors

The continuity equations for electrons and holes in semiconductors are given by [1, pp. 58-63, 2, pp. 62-63]

$$\frac{1}{q}\nabla \cdot \boldsymbol{J}_n + G_n - R_n = \frac{\partial n}{\partial t}, \qquad (2)$$

$$-\frac{1}{q}\nabla \cdot \boldsymbol{J}_{\mathrm{p}} + G_{\mathrm{p}} - R_{\mathrm{p}} = \frac{\partial p}{\partial t}, \tag{3}$$

where $q$ is the elementary charge, $\boldsymbol{J}_{\mathrm{n}}$ ($\boldsymbol{J}_{\mathrm{p}}$) is the electron (hole) current density, $G_{\mathrm{n}}$ ($G_{\mathrm{p}}$) is the electron (hole) generation rate per unit volume due to photon absorption, $R_{\mathrm{n}}$ ($R_{\mathrm{p}}$) is the net electron (hole) recombination rate per unit volume, $n$ ($p$) is the density or concentration of free electrons (holes), and $t$ is time. Carrier generation due to external photon absorption is not a source of carrier generation that inherently occurs in semiconductors, and therefore $G_{\mathrm{n}}$ and $G_{\mathrm{p}}$ have their own separate terms in the continuity equations (i.e. are not included in $R_{\mathrm{n}}$ and $R_{\mathrm{p}}$). For the one-dimensional case at steady state, the continuity equations become [2, p. 63]

$$\frac{1}{q}\frac{\partial J_{\mathrm{n}}}{\partial x} + G_{\mathrm{n}} - R_{\mathrm{n}} = 0, \tag{4}$$

$$-\frac{1}{q}\frac{\partial J_{\mathrm{p}}}{\partial x} + G_{\mathrm{p}} - R_{\mathrm{p}} = 0. \tag{5}$$

The electron current density $J_{\mathrm{n}}$ and the hole current density $J_{\mathrm{p}}$ for the one-dimensional case are given by

$$J_{\mathrm{n}} = q\mu_{\mathrm{n}} F n + q D_{\mathrm{n}} \frac{dn}{dx}, \tag{6}$$

$$J_{\mathrm{p}} = q\mu_{\mathrm{p}} F p - q D_{\mathrm{p}} \frac{dp}{dx}, \tag{7}$$

where $\mu_{\mathrm{n}}$ ($\mu_{\mathrm{p}}$) is the electron (hole) mobility, $F$ is the electric field, $D_{\mathrm{n}} = \mu_{\mathrm{n}} k_{\mathrm{B}} T / q$ is the electron diffusion coefficient, and $D_{\mathrm{p}} = \mu_{\mathrm{p}} k_{\mathrm{B}} T / q$ is the hole diffusion coefficient, where $k_{\mathrm{B}}$ is the Boltzmann constant and $T$ is absolute temperature. Here, we consider the case where there is no temperature gradient (i.e., $dT/dx = 0$) since the effect of the temperature gradient is negligible in general semiconductor devices (hence the current densities consist of the drift and the diffusion components only), unlike in thermoelectric generators where the temperature gradient is purposely maintained.

In general, there are three main recombination mechanisms, which are the band-to-band (also called radiative) recombination, Shockley-Read-Hall (SRH) or trap-assisted recombination, and Auger recombination [1, pp. 39-44]. The total band-to-band recombination rate is proportional to the product of the electron and the hole concentrations (i.e., $\propto np$) [1, pp. 313-315]. The net band-to-band recombination rate per unit volume is therefore [1, pp. 313-315]

$$R_{bb} = k_{bb}np - G_{0(bb)}, \qquad (8)$$

where $k_{bb}$ is the band-to-band recombination coefficient and $G_{0(bb)}$ is the $G_0$ component [see Eq. (1)] for the band-to-band recombination process. It is worth noting that if $R_{bb}$ is zero, it does not necessarily mean that the band-to-band recombination process does not occur, but it simply means that the total band-to-band recombination rate per unit volume (i.e. $k_{bb}np$) is equal to $G_{0(bb)}$. We may write Eq. (8) as [1, pp. 313-315]

$$R_{bb} = k_{bb}\left(np - n_{0(bb)}p_{0(bb)}\right), \qquad (9)$$

which means $G_{0(bb)} = k_{bb}n_{0(bb)}p_{0(bb)}$.

The total SRH recombination rate is also proportional to $np$ [1, pp. 317-324]. Therefore, the net SRH recombination rate per unit volume is given by [1, pp. 317-324]

$$R_{SRH} = k_{SRH}np - G_{0(SRH)}, \qquad (10)$$

where $k_{SRH}$ is the SRH recombination coefficient (see Ref. [1, pp. 317-324] for the detail of $k_{SRH}$) and $G_{0(SRH)}$ is the $G_0$ component [see Eq. (1)] for the SRH recombination process. We can write Eq. (10) as [1, pp. 317-324]

$$R_{SRH} = k_{SRH}\left(np - n_{0(SRH)}p_{0(SRH)}\right), \qquad (11)$$

which means $G_{0(SRH)} = k_{SRH} n_{0(SRH)} p_{0(SRH)}$. Note that since $k_{SRH}$ depends on $n$ and $p$ [1, pp. 317-324], then $G_{0(SRH)}$ will definitely change if $G_n$ and $G_p$ change, as opposed to $G_{0(bb)}$ which could remain constant even if $G_n$ and $G_p$ change.

For the Auger recombination, the total rate is proportional to $n^2 p + p^2 n$ [1, pp. 42-44]. Therefore, the net Auger recombination rate per unit volume is [1, pp. 42-44]

$$R_A = k_A \left(n^2 p + p^2 n\right) - G_{0(A)}, \tag{12}$$

where $k_A$ is the Auger recombination coefficient and $G_{0(A)}$ is the $G_0$ component for the Auger recombination process. To be consistent with $R_{bb}$ and $R_{SRH}$, we can write Eq. (12) as [1, pp. 42-44]

$$R_A = \left(k_A n + k_A p\right)\left(np - n_{0(A)} p_{0(A)}\right), \tag{13}$$

which means $G_{0(A)} = k_A \left(n + p\right) n_{0(A)} p_{0(A)}$. Note that Eq. (13) indicates that $G_{0(A)}$ also changes if $G_n$ and $G_p$ change.

The net electron and the net hole recombination rates per unit volume (i.e., $R_n$ and $R_p$) are sum of all possible processes, and hence are

$$R_n = R_p = R_{bb} + R_{SRH} + R_A. \tag{14}$$

From Eq. (14), it is also easy to see that the total carrier recombination rate is the sum of $k_{bb} np$ (the total band-to-band recombination rate), $k_{SRH} np$ (the total SRH recombination rate), and $k_A \left(n^2 p + p^2 n\right)$ (the total Auger recombination rate). Furthermore, the total $G_0$ is the sum of $G_{0(bb)}$, $G_{0(SRH)}$, and $G_{0(A)}$. Since the total rate of a given recombination process cannot be less than the inherent carrier generation rate (or $G_0$) for that recombination process, this means $R_{bb} \geq 0$, $R_{SRH} \geq 0$ and $R_A \geq 0$. Therefore, when $R_n = R_p = 0$, this means $R_{bb}$, $R_{SRH}$, and $R_A$ must all be zero since $R_{bb}$, $R_{SRH}$, or $R_A$ cannot be negative.

# 3  Models for quantifying the net carrier recombination rate

In Sec. 3.1, we will rederive the conventional model for quantifying $R_{\text{net}}$. The conventional model is derived based on semiconductors that are not part of semiconductor devices (called isolated semiconductors here). In Sec. 3.2, we will derive a new model for quantifying $R_{\text{net}}$ based on semiconductors that are part of semiconductor devices. The purpose we rederive the conventional $R_{\text{net}}$ model is to assist us in deriving and understanding the newly proposed $R_{\text{net}}$ model.

## 3.1  Model for quantifying the net carrier recombination rate in isolated semiconductors

Here, an isolated semiconductor is referred to a semiconductor, intrinsic or extrinsic (doped), that is not part of a semiconductor device, and hence it is not connected to an external circuit and no current can flow inside it. To determine $G_{0(\text{bb})}$, $G_{0(\text{SRH})}$, and $G_{0(\text{A})}$, we need to determine $n_{0(\text{bb})}$, $p_{0(\text{bb})}$, $n_{0(\text{SRH})}$, $p_{0(\text{SRH})}$, $n_{0(\text{A})}$, and $p_{0(\text{A})}$. Let us consider a special case when $R_{\text{n}}$ and $R_{\text{p}}$ are zero. Let us denote the electron concentration $n$ and the hole concentration $p$ in isolated semiconductors when $R_{\text{n}} = R_{\text{p}} = 0$ as $n_{0(\text{is})}$ and $p_{0(\text{is})}$, respectively. As mentioned in Sec. 2, when $R_{\text{n}}$ and $R_{\text{p}}$ are zero, $R_{\text{bb}}$, $R_{\text{SRH}}$, and $R_{\text{A}}$ must all be zero, and hence according to Eqs. (9), (11), and (13), we must have

$$n_{0(\text{is})} = n_{0(\text{bb})} = n_{0(\text{SRH})} = n_{0(\text{A})}, \tag{15}$$

$$p_{0(\text{is})} = p_{0(\text{bb})} = p_{0(\text{SRH})} = p_{0(\text{A})}. \tag{16}$$

The result given by Eqs. (15) and (16) means that according to Eq. (14), we can now write $R_{\text{n}}$ and $R_{\text{p}}$ in isolated semiconductors as

$$R_{\text{n}} = R_{\text{p}} = k_{\text{total}}\left(np - n_{0(\text{is})} p_{0(\text{is})}\right), \tag{17}$$

where $k_{\text{total}} = k_{\text{bb}} + k_{\text{SRH}} + k_{\text{A}}(n+p)$.

If we can determine $n_{0(\text{is})}$ and $p_{0(\text{is})}$, we can determine $G_{0(\text{bb})}$, $G_{0(\text{SRH})}$, and $G_{0(\text{A})}$. When $R_{\text{n}} = R_{\text{p}} = 0$, the carrier continuity equations at steady state become

$$\frac{1}{q}\frac{\partial J_{\text{n}}}{\partial x} + G_{\text{n}} = 0, \tag{18}$$

$$-\frac{1}{q}\frac{\partial J_{\text{p}}}{\partial x} + G_{\text{p}} = 0. \tag{19}$$

Since there is no current flow (i.e., $J_{\text{n}}$ and $J_{\text{p}}$ must be zero everywhere inside an isolated semiconductor), $\partial J_{\text{n}}/\partial x$ and $\partial J_{\text{p}}/\partial x$ must also be zero, which means Eqs. (18) and (19) can be satisfied only if $G_{\text{n}}$ and $G_{\text{p}}$ are zero. This means that the situation $R_{\text{n}} = R_{\text{p}} = 0$ can only happen in an isolated semiconductor if there is no external photon absorption, which is well known.

To find out $n_{0(\text{is})}$ and $p_{0(\text{is})}$, we simply need to solve $J_{\text{n}} = 0$ and $J_{\text{p}} = 0$. An internal electric field can exist inside an isolated semiconductor, for example if the isolated semiconductor is attached to another isolated semiconductor with a different Fermi level, or if the isolated semiconductor is non-uniformly doped. Inside an isolated semiconductor, the electric field $F$ is determined by the gradient of the conduction band minimum $E_{\text{c}}$, given by $F = dE_{\text{c}}/qdx$, or by the gradient of the valence band maximum $E_{\text{v}}$, given by $F = dE_{\text{v}}/qdx$ [2, p. 46].

When $F$ is not zero, the conditions $J_{\text{n}} = 0$ [refer Eq. (6)] and $J_{\text{p}} = 0$ [refer Eq. (7)] lead to

$$\frac{dn_{0(\text{is})}}{dx} = -\frac{n_{0(\text{is})}}{k_{\text{B}}T}\frac{dE_{\text{c}}}{dx}, \tag{20}$$

$$\frac{dp_{0(\text{is})}}{dx} = \frac{p_{0(\text{is})}}{k_\text{B}T}\frac{dE_\text{v}}{dx}. \tag{21}$$

When $F=0$ (hence the gradients of $E_\text{c}$ and $E_\text{v}$ are zero), the conditions $J_\text{n}=0$ and $J_\text{p}=0$ lead to

$$\frac{dn_{0(\text{is})}}{dx}=0, \tag{22}$$

$$\frac{dp_{0(\text{is})}}{dx}=0. \tag{23}$$

Equations (20) and (21) are for isolated semiconductors (intrinsic and extrinsic) when there is an electric field inside the semiconductors, whereas Eqs. (22) and (23) are also for isolated semiconductors (intrinsic and extrinsic) but when there is no electric field inside the semiconductors.

First, we will try to determine $n_{0(\text{is})}$ and $p_{0(\text{is})}$ for the intrinsic semiconductor case. We know that $n_{0(\text{is})}$ and $p_{0(\text{is})}$ are the electron and the hole concentrations, respectively, when the net carrier recombination rate and the external photon absorption are zero. For an isolated semiconductor in general (i.e., intrinsic and extrinsic), no net carrier recombination and no external photon absorption mean the semiconductor is in thermodynamic equilibrium [1, 2]. For an isolated intrinsic semiconductor in thermodynamic equilibrium, the electron concentration and the hole concentration are called the intrinsic electron concentration $n_\text{int}$ and the intrinsic hole concentration $p_\text{int}$, respectively, which are given by [2, p. 20]

$$n_\text{int} = N_\text{c}\exp\left[\frac{-(E_\text{c}-E_\text{Fi})}{k_\text{B}T}\right], \tag{24}$$

$$p_\text{int} = N_\text{v}\exp\left[\frac{-(E_\text{Fi}-E_\text{v})}{k_\text{B}T}\right], \tag{25}$$

where $N_c$ ($N_v$) is the effective density of states in the conduction (valence) band, $E_c$ ($E_v$) is the conduction (valence) band minimum (maximum), and $E_{Fi}$ is the Fermi level of the intrinsic semiconductor. Note that the gradients of $E_c$ and $E_v$ in Eqs. (24) and (25) are zero if there is no electric field but are not zero if there is an internal electric field. However, the gradient of $E_{Fi}$ in Eqs. (24) and (25) must always be zero (whether there is an electric field or not) since Eqs. (24) and (25) is for intrinsic semiconductors in thermodynamic equilibrium (the Fermi level of a material in thermodynamic equilibrium must be uniform) [2].

If there is no electric field ($dE_c/dx$ and $dE_v/dx$ are zero), it can be easily shown that $dn_{int}/dx = 0$ and $dp_{int}/dx = 0$, and therefore letting $n_{0(is)} = n_{int}$ and $p_{0(is)} = p_{int}$ can satisfy Eqs. (22) and (23), respectively. If there is an internal electric field ($dE_c/dx$ and $dE_v/dx$ are not zero), letting $n_{0(is)} = n_{int}$ and $p_{0(is)} = p_{int}$ can also satisfy Eqs. (20) and (21), respectively. Therefore, for both cases (when an electric field does and does not exist), it is clear that the parameters $n_{0(is)}$ and $p_{0(is)}$ for an isolated intrinsic semiconductor is given by $n_{int}$ and $p_{int}$, respectively.

Next, to determine $n_{0(is)}$ and $p_{0(is)}$ for the extrinsic semiconductor case, we follow a similar procedure as for the intrinsic case above. First, we need to know the electron and the hole concentrations in isolated extrinsic semiconductors when there are no net carrier recombination and no external photon absorption (i.e., in thermodynamic equilibrium). When an isolated extrinsic semiconductor is in thermodynamic equilibrium, we know the product of the electron and the hole concentrations is still equal to the product of the electron and the hole concentrations before the semiconductor is doped (i.e., $n_{int} p_{int}$) [2, pp. 20-22]. From above, we know that $n_{int}$ and $p_{int}$ can satisfy the conditions given by Eqs. (20), (21), (22), and (23). Therefore, for an isolated extrinsic semiconductor, $n_{0(is)} p_{0(is)}$ for both cases (when an electric field does and does not exist) is still given by $n_{\text{int}} p_{\text{int}}$, where $n_{\text{int}}$ and $p_{\text{int}}$ are the intrinsic electron and the intrinsic hole concentrations, respectively, inside the extrinsic semiconductor before it is doped.

Since $n_{\text{int}} = p_{\text{int}}$ [2, p. 22], this means $n_{0(\text{is})} p_{0(\text{is})} = n_{\text{int}}^2$. Applying the obtained result to Eqs. (9), (11), and (13), the net band-to-band, the net SRH, and the net Auger recombination rates per unit volume in isolated semiconductors (intrinsic and extrinsic) are therefore given by

$$R_{\text{bb}} = k_{\text{bb}} \left( np - n_{\text{int}}^2 \right), \tag{26}$$

$$R_{\text{SRH}} = k_{\text{SRH}} \left( np - n_{\text{int}}^2 \right), \tag{27}$$

$$R_{\text{A}} = \left( k_{\text{A}} n + k_{\text{A}} p \right) \left( np - n_{\text{int}}^2 \right). \tag{28}$$

The results given by Eqs. (26), (27), and (28) are widely known [1, pp. 39-44]. Therefore, the net carrier recombination rate per unit volume $R_{\text{net}}$ in isolated semiconductors can be generally modeled as

$$R_{\text{net}} = k_{\text{r}} \left( np - n_{\text{int}}^2 \right), \tag{29}$$

where $k_{\text{r}}$ is the general coefficient of the considered recombination process. The recombination model given by Eq. (29) is the conventional model that is derived (and hence valid) for modeling $R_{\text{net}}$ in isolated semiconductors, but the model is also adopted for modeling $R_{\text{net}}$ in semiconductor devices. In the next subsection, we will derive the model to describe $R_{\text{net}}$ in semiconductor devices and show that the model is not as described by Eq. (29).

### 3.2 Model for quantifying the net carrier recombination rate in semiconductor devices

To derive the parameters $n_{0(\text{bb})}$, $p_{0(\text{bb})}$, $n_{0(\text{SRH})}$, $p_{0(\text{SRH})}$, $n_{0(\text{A})}$, and $p_{0(\text{A})}$ for semiconductors that are part of semiconductor devices, we follow a similar procedure as for the isolated semiconductor case above. Again, let us consider the special case when $R_{\text{n}}$ and $R_{\text{p}}$ are zero.

Let us denote the electron concentration $n$ and the hole concentration $p$ in the considered semiconducting material of a semiconductor device when $R_n = R_p = 0$ as $n_{0(d)}$ and $p_{0(d)}$, respectively. When $R_n$ and $R_p$ are zero, $R_{bb}$, $R_{SRH}$, and $R_A$ must all be zero, and therefore according to Eqs. (9), (11), and (13), we have

$$n_{0(d)} = n_{0(bb)} = n_{0(SRH)} = n_{0(A)}, \tag{30}$$

$$p_{0(d)} = p_{0(bb)} = p_{0(SRH)} = p_{0(A)}. \tag{31}$$

When $R_n$ and $R_p$ are zero, the continuity equations for the carriers inside the considered semiconducting material become

$$\frac{1}{q}\frac{\partial J_n}{\partial x} + G_n = 0, \tag{32}$$

$$-\frac{1}{q}\frac{\partial J_p}{\partial x} + G_p = 0. \tag{33}$$

An external voltage bias can be applied to a semiconductor device which can result in non-zero $J_n$ and $J_p$. External photon absorption contributes to the excess free carriers. At any given bias condition, $R_n$ and $R_p$ are expected to reach their lowest value (i.e. zero) when the excess free carriers are at the lowest (i.e. when there is no external photon absorption or $G_n = G_p = 0$). Therefore, at any given bias condition, $R_n = R_p = 0$ should happen when $G_n = G_p = 0$. Considering the case when $G_n$ and $G_p$ are zero, Eqs. (32) and (33) become

$$\frac{\partial J_n}{\partial x} = 0, \tag{34}$$

$$\frac{\partial J_p}{\partial x} = 0. \tag{35}$$

Here we show that both $n_{0(is)}$ (for isolated semiconductors) and $n_{0(d)}$ (for semiconductors that are part of devices) must satisfy Eq. (34), and both $p_{0(is)}$ and $p_{0(d)}$ must satisfy Eq. (35).

However, for the isolated semiconductor case, Eqs. (34) and (35) are automatically satisfied by satisfying the fact that $J_n = J_p = 0$ in isolated semiconductors. For the case of semiconductor devices, Eqs. (34) and (35) must be satisfied without violating the fact that $J_n$ and $J_p$ can be non-zero in semiconductor devices. Therefore, it is clear that letting $n_{0(d)} = n_{int}$ [Eq. (24)] and $p_{0(d)} = p_{int}$ [Eq. (25)] would not give us the general solutions to Eqs (34) and (35), respectively, since $n_{int}$ and $p_{int}$ lead to $J_n$ and $J_p$ that are always zero, which is not true in semiconductor devices. To obtain $n_{0(d)}$ and $p_{0(d)}$, Eqs. (34) and (35) need to be solved numerically together with the Poisson's equation (the Poisson's equation is used to determine the electric field $F$ that appears in the expressions for $J_n$ and $J_p$) and the relevant boundary conditions at the edges of the considered semiconducting material.

To obtain the analytical expressions for $n_{0(d)}$ and $p_{0(d)}$, we need to make a simplification by assuming $F$ to be uniform (i.e. independent of $x$). Applying the assumption and substituting Eqs. (6) and (7) into Eqs. (34) and (35), respectively, we have

$$F \frac{dn_{0(d)}}{dx} + \frac{k_B T}{q} \frac{d^2 n_{0(d)}}{dx^2} = 0, \tag{36}$$

$$F \frac{dp_{0(d)}}{dx} - \frac{k_B T}{q} \frac{d^2 p_{0(d)}}{dx^2} = 0. \tag{37}$$

The general solutions to Eqs. (36) and (37) are

$$n_{0(d)} = A_n \exp\left(\frac{-qFx}{k_B T}\right) + B_n, \tag{38}$$

$$p_{0(d)} = A_p \exp\left(\frac{qFx}{k_B T}\right) + B_p, \tag{39}$$

where $A_n$, $B_n$, $A_p$ and $B_p$ are constants that can be obtained by applying the boundary conditions at the edges of the considered semiconducting material. As opposed to $n_{int}^2$ that appears in the conventional model described in the Sec. 3.1, $n_{0(d)}$ and $p_{0(d)}$ depend on $F$, and hence, depend on the applied bias.

It is worth noting that for the special case when $F = 0$, the first terms on the left-hand side of Eqs. (36) and (37) disappear, and the general solutions for this special case are

$$n_{0(d)} = A_n x + B_n, \tag{40}$$

$$p_{0(d)} = A_p x + B_p. \tag{41}$$

Applying the obtained result to Eqs. (9), (11), and (13), the net band-to-band, the net SRH, and the net Auger recombination rates per unit volume in a semiconducting material of a semiconductor device are therefore given by

$$R_{bb} = k_{bb}\left(np - n_{0(d)}p_{0(d)}\right), \tag{42}$$

$$R_{SRH} = k_{SRH}\left(np - n_{0(d)}p_{0(d)}\right), \tag{43}$$

$$R_A = \left(k_A n + k_A p\right)\left(np - n_{0(d)}p_{0(d)}\right). \tag{44}$$

Based on our result here, we can propose that the net carrier recombination rate per unit volume $R_{net}$ in a semiconductor device can be modeled as

$$R_{net} = k_r\left(np - n_{0(d)}p_{0(d)}\right), \tag{45}$$

where $k_r$ is the general coefficient of the considered recombination process, $n_{0(d)}$ is the electron concentration that results from the solution to $\nabla \cdot \boldsymbol{J}_n = 0$ in the considered semiconducting material of the device, and $p_{0(d)}$ is the hole concentration that results from the solution to $\nabla \cdot \boldsymbol{J}_p = 0$ in the considered semiconducting material of the device. Concerning the physical meaning of $n_{0(d)}$ and $p_{0(d)}$, they are simply the electron and the hole concentrations, respectively, that inherently exist in the considered semiconducting material at any given voltage bias when $G_n = G_p = 0$ and $R_n = R_p = 0$. Further explanation regarding $n_{0(d)}$ and $p_{0(d)}$ will be discussed in Sec. 4.2.

## 4 Model validation and discussion

### 4.1 Method

To support the newly proposed model for describing $R_{net}$ in semiconductor devices (as derived in Sec. 3.2) and discuss its significance, we will compare between the use of the proposed model and the use of the conventional $R_{net}$ model (as derived in Sec. 3.1) in obtaining the current-voltage (J-V) characteristic of a specific semiconductor device, namely organic solar cell (OSC). The reason why OSC is used is because the source code for calculating the J-V characteristic of OSCs is available to us, and hence we can modify the recombination model in the code whether to follow the conventional model or the newly proposed model. The OSC considered here consists of an active layer (a blend of an electron accepting material and an electron donating material) sandwiched between two flat electrodes as described in Ref. [3] and as shown in Fig. 1a.

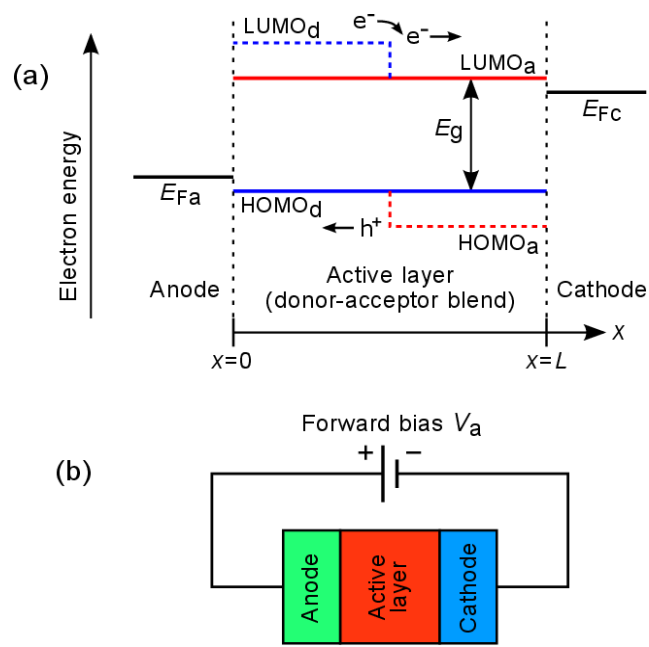

**Fig. 1 a** A schematic of the considered OSC showing the positions and the energy levels of the components. $LUMO_d$ and $HOMO_d$ denote the lowest unoccupied molecular orbital (LUMO) and the highest occupied molecular orbital (HOMO) of the donor material, respectively. $LUMO_a$ and $HOMO_a$ denote the LUMO and the HOMO of the acceptor material, respectively. $E_{Fa}$ and $E_{Fc}$ denote the Fermi levels of the anode and the cathode, respectively. $L$ is the thickness of the active layer. $E_g$ denotes the effective band gap, which is the difference between $LUMO_a$ and $HOMO_d$. **b** Illustration of the forward bias applied voltage $V_a$ on the OSC.

The analytical model developed in Ref. [3] is used here to calculate the J-V characteristics, where the electric field is taken to be uniform and given by

$$F = \frac{V_a - V_{bi}}{L}, \tag{46}$$

where $V_a$ is the applied voltage bias, and $V_{bi} = (E_{Fc} - E_{Fa})/q$ is the built-in voltage. The boundary conditions for the electron concentration $n$ and the hole concentration $p$ at the edges of the active layer (at $x=0$ and $x=L$) in Ref. [3] are given by

$$n\big|_{x=0} = N_c \exp\left[\frac{-(E_g - \varphi_{pa})}{k_B T}\right], \tag{47}$$

$$n\big|_{x=L} = N_c \exp\left(\frac{-\varphi_{nc}}{k_B T}\right), \tag{48}$$

$$p\big|_{x=0} = N_v \exp\left(\frac{-\varphi_{pa}}{k_B T}\right), \tag{49}$$

$$p\big|_{x=L} = N_v \exp\left[\frac{-(E_g - \varphi_{nc})}{k_B T}\right], \tag{50}$$

where $\varphi_{pa}$ is the hole injection barrier at anode given by the difference between $HOMO_d$ and $E_{Fa}$ (refer Fig. 1a), and $\varphi_{nc}$ is the electron injection barrier at cathode given by the difference between $LUMO_a$ and $E_{Fc}$ (refer Fig. 1a).

The model in Ref. [3] considers the bimolecular recombination and the monomolecular recombination. The monomolecular recombination is basically a simplified (approximated) version of the SRH recombination and is neglected in our calculations here. The bimolecular recombination (which is basically the band-to-band recombination) is considered in our calculations here. The net bimolecular recombination rate per unit volume in the active layer according to the conventional $R_{net}$ model (see Sec. 3.1) is

$$R_{b(conv)} = \gamma k_L \left( np - n_{int}^2 \right), \tag{51}$$

where $\gamma$ is the bimolecular recombination reduction coefficient, $k_L = q(\mu_n + \mu_p)/\varepsilon$ is the Langevin recombination coefficient with $\varepsilon$ being the effective permittivity of the active layer, and $n_{int}$ is the intrinsic electron concentration given by Eq. (24). The relationship $E_c - E_{Fi} = E_g/2$ is used to compute $n_{int}$. The net bimolecular recombination rate per unit volume in the active layer according to the new model for describing $R_{net}$ in semiconductor devices (see Sec. 3.2) is

$$R_{b(new)} = \gamma k_L \left( np - n_{0(d)} p_{0(d)} \right), \tag{52}$$

where $n_{0(d)}$ is given by Eq. (38) and $p_{0(d)}$ is given by Eq. (39). For brevity, no calculations at exactly $F = 0$ (i.e., at $V_a = V_{bi}$) will be made in this paper, and hence Eqs. (40) and (41) will not be used here [i.e., only Eqs. (38) and (39) will be used to describe $n_{0(d)}$ and $p_{0(d)}$, respectively]. The constants $A_n$, $B_n$, $A_p$, and $B_p$ for $n_{0(d)}$ and $p_{0(d)}$ can be obtained by applying the boundary conditions (i.e. $n_{0(d)}|_{x=0} = n|_{x=0}$, $n_{0(d)}|_{x=L} = n|_{x=L}$, $p_{0(d)}|_{x=0} = p|_{x=0}$, and $p_{0(d)}|_{x=L} = p|_{x=L}$), which gives

$$A_n = \frac{N_c \exp[-\varphi_{nc}/(k_B T)] - N_c \exp[-(E_g - \varphi_{pa})/(k_B T)]}{\exp[-qFL/(k_B T)] - 1}, \quad (53)$$

$$B_n = N_c \exp[-(E_g - \varphi_{pa})/(k_B T)] - A_n, \quad (54)$$

$$A_p = \frac{N_v \exp[-(E_g - \varphi_{nc})/(k_B T)] - N_v \exp[-\varphi_{pa}/(k_B T)]}{\exp[qFL/(k_B T)] - 1}, \quad (55)$$

$$B_p = N_v \exp[-\varphi_{pa}/(k_B T)] - A_p. \quad (56)$$

The parameter values used in our calculations here are the same as used in Ref. [3] under the 1 sun condition (which are based on the P3HT:PCBM solar cell in Ref. [4]), except for the value of $\gamma$ [refer Eqs. (51) and (52)]. Here, $\gamma = 0.01$ is used compared with $\gamma = 0.001$ used in Ref. [3] and Ref. [4]. A reduction in the bimolecular recombination coefficient in comparison with the Langevin recombination coefficient was observed experimentally [5, 6]. The reduction is partly because the free electrons and the free holes are located in different materials of the active layer (which consists of donor and acceptor networks), and this reduces the probability for them to meet and recombine [7]. $\gamma$ can be affected by the fineness of the donor-acceptor morphology of the interpenetrating networks [8], and can have a value range of $0 \leq \gamma \leq 1$. A higher $\gamma$ is used here compared with the value in Ref. [3] and Ref. [4] because we want to obtain results that can adequately distinguish between the use of the conventional model and the new model.

## 4.2 Results and discussion

Figure 2 illustrates $n_{0(d)}$, $p_{0(d)}$, $n_{int}$, $n_{0(d)}p_{0(d)}$, and $n_{int}^2$ as functions of $x$. It can be seen that $n_{0(d)}p_{0(d)}$ is the same as $n_{int}^2$ at short-circuit point (at $V_a = 0$ V as shown in Fig. 2b) but not at a non-zero bias (e.g. at $V_a = 0.5$ V as also shown in Fig. 2b), which will be explained at the end of this section. As mentioned in Sec. 3.2, unlike $n_{0(d)}$ and $p_{0(d)}$, $n_{int}$ has no dependency on $F$ (and hence on $V_a$), and therefore is invariant when $V_a$ is varied.

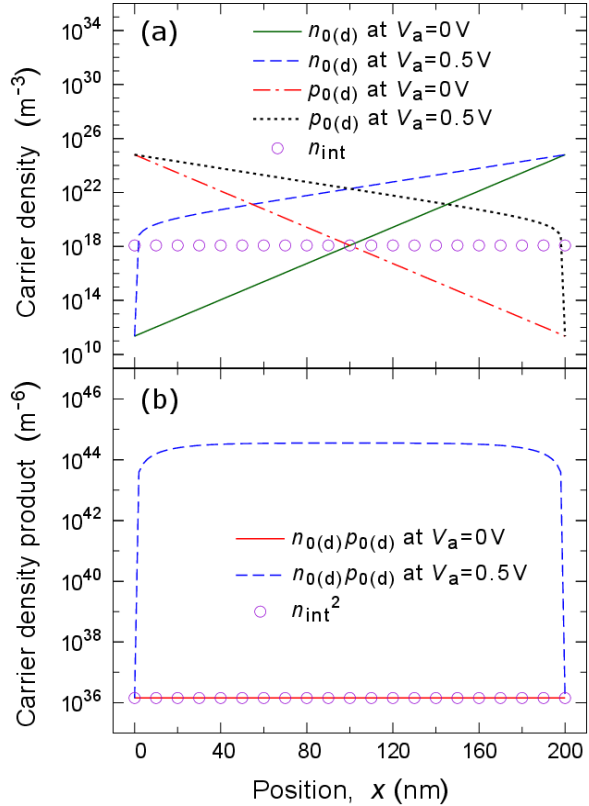

**Fig. 2** Panel **a** shows $n_{int}$ [Eq. (24)] as a function of $x$, and $n_{0(d)}$ [as given by Eqs. (38), (53), and (54)] and $p_{0(d)}$ [as given by Eqs. (39), (55), and (56)] as functions of $x$ at two different applied voltages ($V_a = 0$ V and $V_a = 0.5$ V). Panel **b** shows $n_{int}^2$ as a function of $x$, and $n_{0(d)}p_{0(d)}$ (the product of $n_{0(d)}$ and $p_{0(d)}$) as a function of $x$ at two different applied voltages

($V_a = 0$ V and $V_a = 0.5$ V). All plots are calculated using the parameter values mentioned in Sec. 4.1.

Figure 3 compares the J-V characteristic obtained using the proposed recombination model [using $R_{b(new)}$ given by Eq. (52)] with the J-V characteristic obtained using the conventional recombination model [using $R_{b(conv)}$ given by Eq. (51)]. The J-V characteristic obtained using $R_{b(new)}$ gives the expected J-V trend as $V_a$ is increased, but the J-V characteristic obtained using $R_{b(conv)}$ gives the "wrong" J-V trend as $V_a$ is increased starting from $V_a$ just below $V_{bi}$. The result in Fig. 3 suggests that the newly proposed recombination model is the better model for quantifying the net carrier recombination rate in the semiconducting material of a semiconductor device.

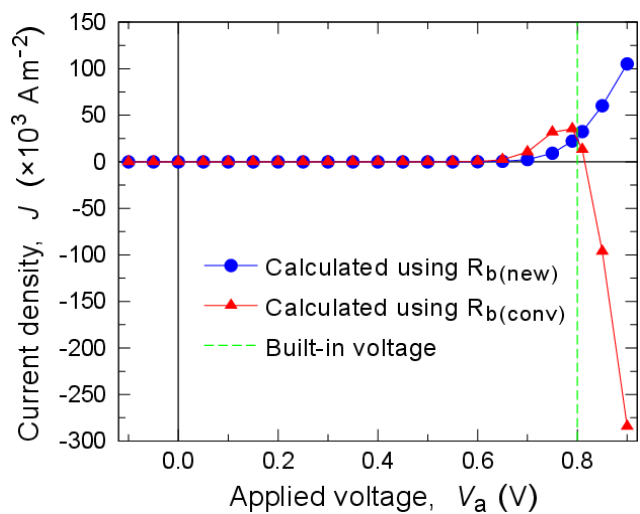

**Fig. 3** The J-V characteristics of the OSC from slightly below the short-circuit to slightly above the built-in voltage $V_{bi}$ ($V_{bi}$ here is 0.8 V, the same as in Ref. [3]). The triangles denote the J-V characteristic obtained using $R_{b(conv)}$ [Eq. (51)] and the circles denote the J-V characteristic obtained using $R_{b(new)}$ [Eq. (52)].

Figure 4 shows that the J-V characteristic obtained using $R_{b(new)}$ is noticeably different compared with the J-V characteristic obtained using $R_{b(conv)}$, where the difference is significant at $V_a$ just below the open-circuit voltage and all the way above that (the difference is due to the increase in the deviation of $n_{0(d)} p_{0(d)}$ from $n_{int}^2$ as $V_a$ is increased away from the short-circuit point). If higher values of $\gamma$ and $k_L$ are used [see Eqs. (51) and (52)], and the SRH recombination is also considered in our calculations, then the difference between the use of the proposed recombination model and the use of the conventional recombination model in predicting the performance of OSCs would be even more significant. Therefore, assuming that the proposed recombination model is the more accurate model for quantifying $R_{net}$ in semiconductor devices (as justified in the derivation in Sec 3.2 and as suggested by the result shown in Fig. 3), we show that the use of the proposed model is important in order to accurately evaluate the performance of OSCs, and we expect that the use of the proposed model is also important to accurately evaluate the performance of other optoelectronic devices whose performance is influenced by carrier recombination such as hybrid perovskite solar cells and LEDs.

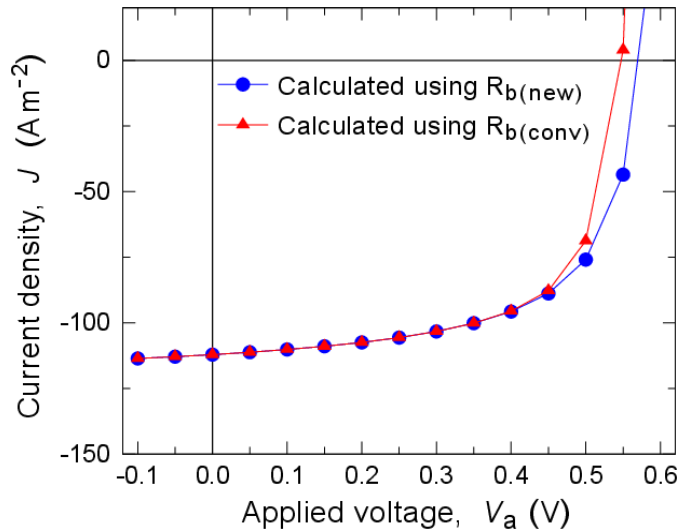

**Fig. 4** The same plot as Fig. 3 but focusing on the J-V characteristics from slightly below the short-circuit to slightly above the open-circuit.

Finally, we will attempt to shed some light on the physical process of $G_0$ [refer Eq. (1)] in the semiconducting material of a semiconductor device. As mentioned earlier in Sec. 1, $G_0$ is the generation rate of free electrons (electrons in the conduction band) and free holes (holes in the valence band) per unit volume that inherently or permanently occurs in a given semiconductor. In isolated semiconductors, it is well known that $G_0$ exists due to thermal agitation [1, pp. 313-324], which can be explained as follows. The product of the free electron and the free hole concentrations that inherently exist in an isolated semiconductor is $n_{int} p_{int}$. This means the total rate for any given recombination process that inherently occurs in an isolated semiconductor is $k_r n_{int} p_{int}$ where $k_r$ is the general coefficient of the considered recombination process [see Eq. (29)]. Since the free electron and the free hole concentrations must be maintained (i.e., the carrier concentrations must not keep increasing or reducing), this means that $G_0$ in an isolated semiconductor must also be the same as the inherent carrier loss rate, which is $k_r n_{int} p_{int}$, and this explains why we obtain $n_{0(is)} p_{0(is)} = n_{int} p_{int}$. It is worth noting that $G_0$ depends on $k_r$ [which may depend on external factors such as photon absorption as discussed in Eq. (11)] since the rate of inherent carrier generation must match the rate of inherent carrier recombination, which may depend on external factors, in order to maintain $n_{int}$ and $p_{int}$. Since $n_{int}$ and $p_{int}$ exist due to thermal excitation [refer Eq. (24) and (25)], then $G_0$ in an isolated semiconductor also simply exist due to carrier thermal excitation (e.g. if $T \to 0$, then $n_{int}$ and $p_{int}$ are basically zero, and this makes $G_0$ in an isolated semiconductor to be zero even if there is photon absorption).

As shown in this paper, the significance of $n_{0(d)} p_{0(d)}$ for semiconductor devices is equivalent to the significance of $n_{int} p_{int}$ (or $n_{int}^2$) for isolated semiconductors. This means $n_{0(d)} p_{0(d)}$ is the product of the free electron and the free hole concentrations that inherently exist in a semiconductor that is part of a semiconductor device at a given voltage bias, where the detail can be explained as follows. As can be seen in Eqs. (38), (39), (53), (54), (55), and (56), carrier injection (represented by the injection barriers and the electric field $F$ that appear in the mentioned equations, where $F$ depends on $V_a$) contributes to $n_{0(d)}$ and $p_{0(d)}$. When a semiconductor device is under a voltage bias, free carriers from outside the semiconducting material of the device can be forced into the semiconductor. Therefore, in the semiconducting material of a semiconductor device, electrons in the conduction band and holes in the valence band can also be created through carrier injection from outside the semiconductor, besides through carrier thermal excitation within the semiconductor (electrons jumping from the valence band to the conduction band in the semiconductor). A semiconductor that is part of a semiconductor device must be connected to an external circuit and its components (otherwise the semiconductor is an isolated semiconductor where no electric current can flow), and therefore, carrier generation due to carrier injection should be considered as part of the carrier generation that inherently occurs in a semiconductor that is part of a device. This means that when a device is under a voltage bias, additional free carriers are injected into the semiconducting material of the device, and therefore, $n_{0(d)} p_{0(d)}$ must be larger than $n_{int} p_{int}$, and this explains the results shown in Fig. 2. Furthermore, when a device is not under an applied bias ($V_a = 0$), no carriers are injected into the semiconducting material of the device, and therefore, $n_{0(d)} p_{0(d)}$ must be the same as $n_{int} p_{int}$, and this is shown in Fig. 2. Since $n_{0(d)}$ and $p_{0(d)}$ exist due to carrier thermal excitation and carrier injection, $G_0$ in the semiconducting material of a semiconductor device also exist due to carrier thermal excitation and carrier injection (e.g., when $V_a = 0$ and $T \to 0$, $G_0$ in a semiconductor device is basically zero even if there is photon absorption).

## 5      Conclusion

We have investigated the modeling for the net carrier recombination rate in semiconductors. The net carrier recombination rate per unit volume $R_{net}$ in a semiconductor is conventionally modeled as $R_{net} = k_r (np - n_{int}^2)$ where $k_r$ is the general coefficient of the considered recombination process, $np$ is the product of the electron and the hole concentrations in the semiconductor, and $n_{int}$ is the intrinsic electron concentration in the semiconductor. The conventional $R_{net}$ model is derived based on a condition that the electric current in the considered semiconductor is always zero, which is valid only if the semiconductor is not part of a semiconductor device. To produce a better and more realistic model for describing $R_{net}$ in semiconductor devices, we have derived and proposed a new model, where the model was derived by considering the fact that electric current can exist in the semiconducting materials of semiconductor devices. We found that the net carrier recombination rate per unit volume $R_{net}$ in a semiconductor that is part of a semiconductor device can be modeled as $R_{net} = k_r (np - n_{0(d)} p_{0(d)})$ where $k_r$ is the general coefficient of the considered recombination process, and $n_{0(d)}$ and $p_{0(d)}$ are carrier concentrations in the semiconductor that result from the solutions to $\nabla \cdot \boldsymbol{J}_n = 0$ and $\nabla \cdot \boldsymbol{J}_p = 0$, respectively, where $\boldsymbol{J}_n$ and $\boldsymbol{J}_p$ are the electron and the hole current densities in the semiconductor, respectively. By using organic solar cells (OSCs) as a case study, we analyzed the calculated J-V characteristics and showed that the newly proposed recombination model is the better model for describing $R_{net}$ in OSCs, and this is expected to be the case for other semiconductor devices too. We showed that the use of the proposed model is important in order to accurately evaluate the performance of OSCs, and therefore we expect that the use of the proposed model is also important to better evaluate the performance of other optoelectronic devices whose performance is influenced by carrier recombination such as LEDs. Finally, we have also provided some physical insights on $n_{0(d)}$ and $p_{0(d)}$. We have explained that $n_{0(d)}$ and $p_{0(d)}$ are the electron concentration in the conduction band and the hole concentration in the valence band, respectively, that inherently exist in the semiconducting material of a semiconductor device at a given voltage bias, where $n_{0(d)}$ and $p_{0(d)}$ can be contributed by the carrier thermal generation from inside the semiconductor as well as by the carrier injection from outside the semiconductor.

**Acknowledgments**

MLII acknowledges the support from the Ministry of Higher Education of Malaysia for the Fundamental Research Grant Scheme (FRGS/1/2017/STG02/UIAM/03/2). Support for AZ work was provided from the Ministry of Education and Science of the Russian Federation (Project 14.Y26.31.0010) and the Russian Science Foundation (№ 19-73-30023), and partial financial support by Welch Foundation grant AT 1617 is also highly appreciated.

**References**

1. S. J. Fonash, Solar Cell Device Physics, 2nd edn. (Academic Press, Massachusetts, 2010)
2. S. M. Sze, K. K. Ng, Physics of Semiconductor Devices, 3rd edn. (John Wiley & Sons, New Jersey, 2007)
3. M. L. Inche Ibrahim, Semicond. Sci. Technol. **33,** 125005 (2018)
4. J. Kniepert, I. Lange, N. J. van der Kaap, L. J. A. Koster, D. Neher, Adv. Energy Mater. **4,** 1301401 (2014)

5. A. Pivrikas, G. Juška, A. J. Mozer, M. Scharber, K. Arlauskas, N. S. Sariciftci, H. Stubb, R. Österbacka, Phys. Rev. Lett. **94,** 176806 (2005)

6. G. Juška, K. Arlauskas, J. Stuchlik, R. Österbacka, J. Non-Cryst. Solids **352,** 1167 (2006)

7. C. M. Proctor, M. Kuik, T. Q. Nguyen, Prog. Polym. Sci. **38,** 1941 (2013)

8. R. G. E. Kimber, A. B. Walker, G. E. Schroder-Turk, D. J. Cleaver, Phys. Chem. Chem. Phys. **12,** 844 (2010)